\documentclass[a4paper,11pt,twoside,onecolumn]{article}

\usepackage{graphicx}
\usepackage{epsf}
\usepackage{amssymb}
\usepackage{sidecap}

\onecolumn
\def\sloppy{\tolerance=100000\hfuzz=\maxdimen \vfuzz=\maxdimen}
\newcommand{\rme}{\mathrm{e}}
\newcommand{\rmd}{\mathrm{d}}
\newcommand{\rms}{{\{d\}}}
\newcommand{\cN}{\mathcal{N}}
\newcommand{\rMFA}{\mathrm{MFA}}
\newcommand{\bfr}{\mathbf{r}}

\vspace{-8ex}\date{}

\begin{document}
\sloppy

\title{\large\bf STATISTICS EFFECTS IN EXTREMAL BLACK HOLES ENSEMBLE}

\author{A.M. Gavrilik${}^\dag$ and A.V. Nazarenko${}^\ddag$\\
{\small\it Bogolyubov Institute for Theoretical Physics of NAS of Ukraine,}\\
{\small\it 14-b, Metrolohichna str., UA--03143 Kyiv, Ukraine}\\
{\small ${}^\dag$omgavr@bitp.kiev.ua, ${}^\ddag$nazarenko@bitp.kiev.ua}}

\maketitle

\begin{abstract}
We consider the grand canonical ensemble of the static and extremal black holes,
when the equivalence of the electric charge and mass of individual black hole is
postulated. Assuming uniform distribution of black holes in space, we are
finding the effective mass of test particle and mean time dilation at
the admissible points of space, taking into account the gravitational action
of surrounding black holes.
Having specified the statistics that governs extremal black holes, we study its effect
on those quantities. Here, the role of statistics is to assign a statistical weight
to the configurations of certain fixed number of black holes.
We borrow these weights from Bose--Einstein, Fermi--Dirac, classical and infinite statistics.
Using mean field approximation, the aforementioned characteristics are
calculated and visualized, what permits us to draw the conclusions on
visible effect of each statistics.


\end{abstract}

\section{Introduction}

Due to the extremality property consisting in the equivalence
of electric charge and mass of a single black hole,\footnote{It
is expected that the extremal configuration for a charged black hole
is the endpoint of Hawking evaporation and corresponds to stable
quantum groundstate~\cite{H75,GS92}.} it became
possible to find a static solution to the Einstein--Maxwell
equations for ensemble of many identical black holes~\cite{Maj,Pap,HH72}
and to study unique physical phenomena under various
conditions.
Quantum mechanical view on a system of such ''particles''
has led to a natural question about the type of statistics
governing the collection of extremal black holes~(BHs)~\cite{Strom93}.
Discussing this, a number of models has appeared
\cite{Strom93,Med96,DK16,Dil17},
using a wide class of deformed (quantum) algebras which
generalize the Heisenberg algebra. An equally important issue
remains to reveal the statistics manifestation
in meaningful effects. Discarding possible high-energy
processes involving black holes, their ensemble is often
associated with a quantum gas \cite{Med96,Dil17,SK18}.
Such a treatment is not always physically correct. This
motivated us to try out a heuristic approach in order to analyze
the properties of the static black hole system without violating
the physical basics of its equilibrium existence.

Here we are not dealing with any aspect that concerns quantum gravity,
including BH statistics inferred therefrom. We do not also appeal to thermodynamics
of (macroscopic) single black hole with large number of degrees of freedom and
the effects of Hawking radiation.

In this paper we limit ourselves by studying the macroscopic properties of
the ensemble formed of the classical BH static configurations, which are most
probable in the specified statistics. This is somewhat similar to a system
of spins that can change their direction in lattice sites, but are not
moving in space. Here, considering the gravitating objects, we are interested
in a mean value of the time scale factor of space-time metric, related to
the gravitational potential, in contrast to the magnetic field
of the spin system. It is expected that the required dependence
of the averaged interaction potential on the mean number of
black holes will be regulated by the statistics used.

From the point of view of the general relativity,
our goal is to evaluate the average energy-mass of the test
particle and the mean time dilation. While the first
characteristic may depend on both the gravity
and statistics, the second is a property of space-time.
Here we examine (quantum) statistics of four types, including the infinite statistics
introduced in \cite{Gr90,Moh90} and exploited in \cite{Strom93}.
We consider it is important to make a comparison of macroscopic
quantities found within the Bose--Einstein, Fermi--Dirac,
classical and infinite statistics.

In the next Section, we statistically describe a grand
canonical ensemble of extremal black holes, In the Section~3 we find
characteristics for four types of statistics. We finish
the paper with discussion and conclusions.

\section{Statistical Characteristics of Black Holes Ensemble}

We start from Majumdar-Papapetrou (MP) solution \cite{Maj,Pap} of the Einstein-Maxwell
equations for $N$ static and extremal black holes (BHs) with equal masses $m$ and
electric charges $Q=m$ (in units $G=c=\hbar=1$), centered at points
$\{{\bf a}_i\in\mathbb{R}^3\}^N_{i=1}$:
\begin{eqnarray}
&&\rmd s^2=-U^{-2}_N(\bfr)\,\rmd t^2+U^2_N(\bfr)\,\rmd\bfr^2,\quad
U_N({\bf r})=1+\sum\limits_{i=1}^N\frac{m}{|{\bf r}-{\bf a}_i|};\label{MP}\\
&&\rmd\bfr^2=\rmd r^2+r^2(\rmd\theta^2+\sin^2{\theta}\rmd\phi^2).\nonumber
\end{eqnarray}
Expressions (\ref{MP}) determine space-time metric $g_{\mu\nu}(\bfr)$
and electrostatic field potential $A_t=U^{-1}_N(\bfr)$ in
$\mathcal{M}_4\simeq\mathbb{R}\times(\mathbb{R}^3\backslash \{B_i\}^N_{i=1})$,
where $B_i=\{\bfr\in\mathbb{R}^3|\, |\bfr-\mathbf{a}_i|\leq r_+\}$
is a ball of radius $r_+=m$, corresponded to the event horizon of extremal
black hole. In the case of a single BH centered at $\mathbf{a}_1=\mathbf{0}$,
replacement $r\to r-m$ leads to the Reissner-Nordstr\"om solution~\cite{HH72}.

Here, we are interested in heuristic study of the macroscopic properties of
BH ensemble by considering different statistics including the infinite
one, argued in \cite{Strom93}. In the case of the static and diagonal
metric (\ref{MP}), it seems appropriate to find a mean value of
$\sqrt{-g_{tt}}$, giving us both the gravitational and electrostatic
potential $A_t$, generated by the extremal BH system. Moreover, it
measures the gravitational time dilation as explained below.

Considering the rest reference frame, let the $N$ randomly distributed (in space)
extremal BHs create background field (geometry) at a given position ${\bf r}$
of static test particle. Accordingly to (\ref{MP}), the infinitesimal
proper time interval is equal there to $\rmd\tau(\bfr)=U^{-1}_N(\bfr)\rmd t$,
where $t$ is a global time (the use of which makes the metric tensor
static\cite{LL}). Then, a local time dilation effect is evaluated by the ratio of proper
time and global time intervals $\Delta\tau/\Delta t$, determined by function
$U^{-1}_N(\bfr)$. Thus, the proper time flows more slowly with
increasing $U_N(\bfr)$.

Note that the equation $U^{-1}_N(\bfr)=\mathrm{const}$ generates equipotential
orbits (closed surfaces) forming, in general, the multiply-connected domain
of equal time flow.

Further, we turn to the relativistic mechanics of a particle with
mass $m_0$, measured at an infinitely large distance from the sources of
the gravitational field. When freely moving, the particle energy $E$ in a static
field is conserved~\cite{LL} and defined as
\begin{eqnarray}
E&=&-m_0\, g_{tt}\,\frac{\rmd t}{\rmd s}=m_0\,\frac{\sqrt{-g_{tt}}}{\sqrt{1-v^2}};\\
v&=&\frac{\sqrt{g_{ij}\,\rmd x^i\rmd x^j}}{\rmd\tau}=\frac{\sqrt{g_{ij}\,\rmd x^i\rmd x^j}}{\sqrt{-g_{tt}}\,\rmd t},
\end{eqnarray}
where $g_{ij}$ are the spatial components of $g_{\mu\nu}$.

Substituting $g_{tt}=U^{-2}_N(\bfr)$ into $E$, we also impose the kinematic constraint
$v=0$ to exclude the relativistic time dilation. It leads to the energy-mass of a static
probe particle in space-time, influenced by the $N$ BHs:
\begin{equation}\label{EN}
E_N(\bfr)=m_0\,U^{-1}_N(\bfr),\qquad
E_0=m_0.
\end{equation}

Note that, comparing $E_N(\bfr)$ and electrostatic potential $A_t=U^{-1}_N(\bfr)$,
$E_N(\bfr)$ may be interpreted as the energy of attraction of the charged probe to
the set of $N$ identical, but opposite charges (being in equilibrium due to gravity).

Let a (local) statistical weight of BH configuration be described by
the Gibbs measure $\exp{(-\beta E_N(\bfr))}$, controlled by an effective
temperature $T=\beta^{-1}$. Dimensionless parameter $x=\beta m_0$, sometimes
used instead of temperature characteristic, is actually of order of magnitude
$m_0/m$ and supposed to be small.

Since the general properties are mostly encoded in the mass, we expect
the energy-mass variations due to collective interaction and quantum
statistics effects.
Thus, we are finding an effective mass $m_*(\bfr)$ of a test particle
in the grand canonical ensemble of extremal black holes. In other words,
we are deriving an interaction potential in terms of macroscopic
(thermodynamic) parameters, the attractive property of which should lead
to $m_*/m_0\leq1$ at $m_0\to0$.

It is worth to note Refs.~\cite{Dil17,SK18}, where the BH mass change is achieved
by modifying the Einstein equations because of using a deformed statistics.
Estimation how the black hole could be affected by the other black hole in a
binary system is given in \cite{SB16}. Note also Refs.~\cite{Bon93,Bini07},
demonstrating an equilibrium between a probe particle and a particle source of
(static) gravitational and electric field, if these are extremal and have both
equal masses and electric charges.

Formulating a problem in $(z,T,V)$ terms, a grand partition function is
\begin{eqnarray}
\mathcal{Z}_\rms(\bfr)&=&\sum\limits_{N=0}^\infty d_N\, z^N\, Q_N(\bfr),
\label{GF}\\
Q_N(\bfr)&=&\int\left(\prod\limits_{i=1}^N\rmd\mathbf{a}_i\right)
\rho_N(\{\mathbf{a}_i\})\exp{(-\beta E_N(\bfr))}.
\end{eqnarray}
Here $z$ is the fugacity; $Q_N$ is the configuration integral;
$\rho_N(\{\mathbf{a}_i\})$ gives a distribution of BHs in space;
$d_N$ is a combinatorial or degeneracy factor ($d_{N=0}=1$),
determined by statistics. Particularly, $d_N=1$ is for usual
Bose-like systems (denoted by ``BE''); $d_N=1/N!$
corresponds to the classical (Cl) systems of identical particles.
In the general case of undetermined statistics, we assign the index
``$\rms$'' to $\mathcal{Z}_\rms$ and other functions.
Note  that in \cite{Min97,BMS01} it was demonstrated how to use
different constructions for the partition function in case of some black holes.

In the static model, no work is performed, and thermodynamics
looks rather restricted. Therefore, we are mainly focusing on finding
the effective mass
\begin{equation}\label{mdef}
m_*(\bfr)=-(\partial_\beta \ln{\mathcal{Z}_\rms(\bfr)})_{z,V}.
\end{equation}
There is also a possibility of determining a mean of
$m_*(\bfr)$ in space. However, such a procedure is not used
here because of the following assumptions.

Limiting ourselves by considering a homogeneous distribution of BHs,
the function $\rho_N(\{\mathbf{a}_i\})$ is supposed to be a constant.
When $\sum_{i=1}^N\mathbf{a}_i\simeq\mathbf{0}$, we can
concentrate on calculating $\mathcal{Z}_\rms=\mathcal{Z}_\rms(\mathbf{0})$
due to the translation invariance inside volume $V$, containing
the BHs, and neglecting the surface effects.

In general case, we require a BH interior inaccessibility for
a probe i.e., $|\bfr-\mathbf{a}_i|>r_+=m$ for all $\mathbf{a}_i$,
and a smallness of $\mathrm{Vol}(B_i)=4\pi m^3/3$ in comparison
with the total volume $V=4\pi R^3/3$, $R>m$.

Taking these requirements into account, we put
\begin{equation}
\rho_N(\{\mathbf{a}_i\})=[(1-\mu^3)V]^{-N},\quad \mu\equiv\frac{m}{R},
\end{equation}
but the overlapping black holes horizons are not mathematically
forbidden here for the sake of calculations simplicity.

The parameter $\mu$ ($0\leq\mu\leq1$) is a measure of gravitational interaction
within the spherical domain of radius $R\sim V^{1/3}$, and it can be regarded
as independent dimensionless model parameter (instead of $V$).

At the first sight, partition function $\mathcal{Z}_\rms$ takes into account
the arbitrarily large number of finite-size balls $\{B_i\}_{i=1}^{N\to\infty}$
within the finite domain. Actually, the mean number $\cN_\rms$ of BHs
within the ensemble is
\begin{equation}\label{N1}
\cN_\rms=(z\partial_z \ln{\mathcal{Z}_\rms})_{T,V},
\end{equation}
and is consistent with $V$ by $\mu^{-3}>\cN_\rms$.

After the assumptions made, we resume that the properties of the black hole
system are determined by the fugacity $z$ and the parameter $\mu$, which fixes the size
(volume) of the system if the mass $m$ of BHs is given. Although the latter means the
presence of a boundary, with a uniform distribution of BHs (as particles of finite size),
the number of BHs is regulated by $z$ and the specified statistics. At the same time,
the temperature controls the effect of the BH system on the probe particle.

Being interested in observables that are extracted by means of a test particle at different
points in the bulk, we are actually studying an ensemble of static test particles (whose
formulation is simplified in the homogeneous case) against a background of the system of BHs.
The Gibbs distribution shows the impossibility to extract any information about the extremal
black hole system at $x\to\infty$ and allows us to smooth out fluctuating quantities.

\textbf{Partition Function Calculation.}
Averaging over black hole configurations, we use an auxiliary formula~\cite{PBM}:
\begin{equation}\label{trans}
\rme^{-1/\kappa}=1-\int_0^\infty J_1(p)\, \rme^{-p^2\kappa/4}\, \rmd p,\qquad \kappa>0,
\end{equation}
where $J_1(p)$ is the Bessel function of the first kind.

Due to translational invariance, integral $Q_N$ takes the form:
\begin{eqnarray}
Q_N&=&1-\int_0^\infty J_1(p)\,\rme^{-p^2/(4x)}(\xi(p))^N\,\rmd p,\quad
x=\beta m_0,\label{trans2}\\
\xi(p)&=&\frac{1}{(1-\mu^3)V}\int\exp{\left(-\frac{p^2}{4x}\frac{m}{|\mathbf{a}|}\right)}
\rmd\mathbf{a}.
\end{eqnarray}
One has
\begin{eqnarray}
(1-\mu^3)\,\xi(p)&=&\frac{3}{R^3}\int_m^R\exp{\left(-\frac{p^2}{4x}\frac{m}{a}\right)}a^2\rmd a\\
&=&\left(2-s\mu+s^2\mu^2\right)\frac{\rme^{-s\mu}}{2}
-\frac{\mu^3}{2}\left[(2-s+s^2)\rme^{-s}\right.\nonumber\\
&&\left.+s^3E_1(s\mu)-s^3E_1(s)\right],\label{x2}
\end{eqnarray}
where $s=p^2/(4x)$, and
\begin{equation}
E_n(s)=\int_1^\infty \rme^{-ts} t^{-n}\rmd t
\end{equation}
is the exponential integral.

Behavior of $\xi$ is mostly determined by $\exp{(-s\mu)}$ and is
corrected by means of expansion:
\begin{equation}
(1-\mu^3)\rme^{s\mu} \xi=(1-\mu^3)-\left(2\mu^4-3\mu^3+\mu\right)\frac{s}{2}+O(s^2),
\end{equation}
where the coefficient by $s^n$ for any $n\geq0$ vanishes at $\mu=1$.

Accounting only for the first term in the r.h.s., we approximate $\xi$
by the function, reproducing its leading properties:
\begin{equation}
\tilde\xi(p)=\exp{\left(-\frac{p^2}{4x}\mu\right)},\quad
\tilde\xi(0)=\xi(0),\quad
\tilde\xi(p)\stackrel{p\to\infty}{\longrightarrow}\xi(p).
\end{equation}

Substituting $\tilde\xi$ instead of $\xi$ into (\ref{trans2}), we arrive at
\begin{equation}\label{EN2}
{\tilde Q}_N=\exp{\left(-\frac{x}{1+\mu N}\right)}.
\end{equation}

Note that the expression for ${\tilde Q}_N$ can be formally
obtained in another ways. Indeed, one of these consists in naive
replacement of $U_N(\mathbf{0})$ with ${\tilde U}_N=1+mN/R_h$,
where $R_h=N/(\sum_{i=1}^N |\mathbf{a}_i|^{-1})$ is a harmonic mean
distance, which does not require a further averaging over
$\{\mathbf{a}_i\}$.

Another way suggests to use the multipole expansion of
$U_N(\mathbf{R})$, when $R\gg|\mathbf{a}_i|$ for all $\mathbf{a}_i$,
which is limited by the first term to give ${\tilde U}_N=1+mN/R$.
Thus, these examples justify the form of ${\tilde Q}_N$
in the leading order approximation. However, the corrections
to ${\tilde Q}_N$ in the next to leading approximation would differ.

The grand partition function is therefore reduced to
\begin{equation}\label{pf1}
{\tilde{\mathcal{Z}}}_\rms=\sum\limits_{N=0}^\infty d_N\,z^N \exp{\left(-\frac{x}{1+\mu N}\right)}.
\end{equation}
In particular, one has ${\tilde{\mathcal{Z}}}_\rms(\mu=0)=\psi_\rms(z)\rme^{-x}$, where
\begin{equation}
\psi_\rms(z)=\sum\limits_{N=0}^\infty d_N\,z^N.
\end{equation}
It leads to $m_*=m_0$ as it must be, when $R\to\infty$ ($\mu\to0$).

Function ${\tilde{\mathcal{Z}}}_\rms$ is of independent interest and may be applicable,
for instance, to a study of the quantum Bose-like many-particle systems with
the $\mu$-deformed spectrum~\cite{Jan93,ChGN}, proportional to $[N]_\mu=N/(1+\mu N)$
and accounting effectively for attraction. Expression (\ref{pf1}) is also suitable
for constructing the models which use the different statistics by choosing $d_N$.

Note that the probability distribution function for a one-dimensional random
walk (RW) along axis $y\in(-\infty;+\infty)$ in time $t>0$ with a drift velocity~$v$,
\begin{equation}
{\cal P}(y,\tau;v)=\frac{1}{\sqrt{4\pi t}}\exp{\left(-\frac{(y-v\tau)^2}{4\tau}\right)},
\end{equation}
can be formally used to re-write (\ref{pf1}) as
\begin{equation}
{\tilde{\cal Z}}_\rms=\rme^{-y^2/4}
\sum\limits_{N=0}^\infty d_N \sqrt{\tau_N}\,
\frac{{\cal P}(y,\tau_N;v)}{{\cal P}(y,\tau_0;v)}
\end{equation}
at $y=2\sqrt{x}$ and $v=2\sqrt{-\mu^{-1}\ln{z}}$; the presence of $v\not=0$ is
crucial for convergence. Thus, ${\tilde{\cal Z}}_\rms$ determines a mean
of $\sqrt{\tau_N}$ with discrete evolution parameter $\tau_N=1+\mu N$ and
$\tau_0=1$ in RW model. Although this correspondence results from the form of
(\ref{MP}), application of RW model to the BH system is justified by a random
distribution of BHs: inclusion of each new BH into ensemble corresponds
to a time step of RW.

Using (\ref{pf1}), let us introduce a mean field (see \cite{ChGN}):
\begin{equation}\label{eta1}
\sigma_\rms(z,x,\mu)=\frac{1}{{\tilde{\cal Z}}_\rms}
\sum\limits_{N=0}^\infty \frac{d_N\,z^N}{1+\mu N} \exp{\left(-\frac{x}{1+\mu N}\right)},
\end{equation}
which immediately defines the effective mass $m_*=m_0\,\sigma_\rms$
due to (\ref{mdef}). It is independent on space and can be evaluated
numerically.

Similarly, we determine the mean number of BHs (see (\ref{N1})):
\begin{equation}\label{Nd}
\cN_\rms(z,x,\mu)=\frac{1}{{\tilde{\cal Z}}_\rms}
\sum\limits_{N=0}^\infty N\,d_N\,z^N \exp{\left(-\frac{x}{1+\mu N}\right)}.
\end{equation}

The field $\sigma_\rms$  also characterizes a mean time dilation
$\langle\Delta\tau/\Delta t\rangle$
within the BH ensemble obeying statistics, denoted as ``$\rms$''.
An influence of the specified statistics on energy-mass and
time dilation consists in non-equal accounting for different BH
configurations by determining weight coefficients $d_N$.
In particular, omitting the fugacity $z$ at the moment,
we see that the BE statistics ($d_N=1$) provides equal
weights for all configurations, while the Cl statistics
($d_N=1/N!$) suppresses the many-BH configurations at the same
macro parameters.

To evaluate $\sigma_\rms$ analytically, we replace the quotient
$1/(1+\mu N)$ in (\ref{pf1}) with expression $\sigma+s_N$,
where $0\leq\sigma\leq1$ is an arbitrary constant (mean field) at the moment,
and $s_N=1/(1+\mu N)-\sigma$ plays the role of fluctuation. Expanding
${\tilde{\cal Z}}_\rms$ over $s_N$ we arrive at
\begin{equation}
{\tilde{\cal Z}}_\rms=\sum\limits_{N=0}^\infty d_N\,z^N \rme^{-x\sigma}
\left[1+\sum\limits_{k=1}^\infty\frac{(-x)^k}{k!}(s_N)^k\right],
\end{equation}
At this stage, ${\tilde{\cal Z}}_\rms$ does not depend on $\sigma$. Also
smallness of $x$ is assumed.

We fix $\sigma$ from the condition $\sum_{N=0}^\infty d_N\,z^N s_N=0$.
It gives us that
\begin{equation}\label{sig}
\sigma=\Lambda_\rms(z,\mu),\qquad
\Lambda_\rms(z,\mu)=\frac{\Phi_\rms\left(z,1,\mu^{-1}\right)}{\mu\, \psi_\rms(z)},
\end{equation}
where
\begin{equation}
\Phi_\rms(z,s,a)=\sum_{N=0}^\infty \frac{d_N\, z^N}{(a+N)^s}.
\end{equation}

Accordingly to the rules of a mean field approximation (MFA), we have
\begin{equation}\label{pf2}
\mathcal{Z}^\rMFA_\rms=\psi_\rms(z)\,\rme^{-x\sigma},
\end{equation}
where $\sigma$ is regarded as independent parameter by
calculating the thermodynamic relations (derivatives). The constraint
$\sigma=\Lambda_\rms(z,\mu)$ should be substituted into the final expressions only.

Substituting (\ref{pf2}) instead of ${\cal Z}_d$ into (\ref{mdef}),
we come to
\begin{equation}\label{eta2}
\sigma^\rMFA_\rms=\Lambda_\rms(z,\mu).
\end{equation}
Thus, we see that
\begin{equation}
\sigma^\rMFA_\rms(z,\mu)=\sigma_\rms(z,0,\mu),
\end{equation}
what follows from (\ref{eta1}).

The mean number of BHs is easily found from (\ref{N1}) to give
\begin{equation}\label{N2}
\cN^\rMFA_\rms(z)=z\partial_z\ln{\psi_\rms(z)}.
\end{equation}

Note that (\ref{eta2}) and (\ref{N2}) represent the extremal BH ensemble
characteristics at the vanishing probe mass $m_0\to0$ and allow us to determine
both a ratio $m_*/m_0$ and a mean time dilation for a specified statistics.

In principle, function $\sigma_\rms(\cN)$ characterizes rather
approximately an effective energy-mass $m_*$ of a probe particle
because of neglecting the other kinds of interaction besides of
the gravitational one. However, $\sigma_\rms(\cN)$ at $m_0\to0$
describes, independently of a probe presence, two significant effects:
1) a mean time dilation as a property of space-time itself, and
2) an influence of the statistics that distinguishes statistically
significant BH configurations, which further form collectively
the properties of space-time.

\section{Statistics Specified}

For a better understanding, let us recall main ideas of our approach. Actually,
we intend to calculate a number of average and universal characteristics,
the numerical values of which should be determined by the BH parameters,
BHs distribution in space and statistical weights dictated by statistics for
a given number of BHs. That means the temperature and the test mass are playing
an auxiliary role and should be excluded at this stage from description.

Although there are several ways to calculate averages, we follow the statistical physics of
systems with an arbitrary number of particles. Writing down the partition function depending
formally on $T$ and $m_0$, we derive the required averages from it, using standard thermodynamic
relations. {\it After} the calculation of derivatives, the restriction $m_0=x=0$ should be applied.
Then the label ``MFA'' means a number of transformations that allow to extract the values of
quantities at $x=0$ and to find corrections to them at $x\not=0$, if needed.

Analyzing the formulas (\ref{eta2}) and (\ref{N2}), we can see that these functions
represent the weighted arithmetic means. This is not surprising, since the energy and number
of particles are additive quantities.

\textbf{Bose--Einstein Statistics.}
To test our approach and to analyze its outcomes, let us first consider
a simplest case of the Bose-like statistics, when the degeneracy factors
are $d_N=1$. Then, the auxiliary functions defined at $0\leq z<1$ read
\begin{equation}
\psi_\mathrm{BE}(z)=\frac{1}{1-z},\quad
\Lambda_\mathrm{BE}(z,\mu)=\frac{1-z}{\mu}\,\Phi\left(z,1,\mu^{-1}\right),
\end{equation}
where $\Phi(z,s,a)=\sum_{N=0}^\infty z^N\,(a+N)^{-s}$ is the Lerch transcendent.

\begin{SCfigure}[1.0]
    \includegraphics[width=6.2cm]{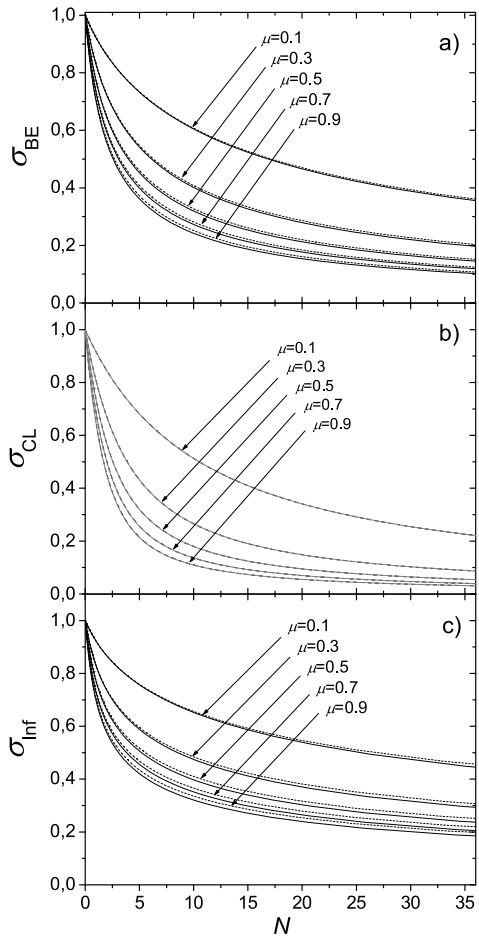}
  \caption{\small Mean field $\sigma_\rms$ versus the BH mean number $\cN$
at different $\mu$ within: a) Bose-like statistics,
b) classical statistics, c) infinite statistics.
Solid curves represent the dependence of $\sigma_\rms$ on $\cN_\rms$,
given parametrically by (\ref{eta1}), (\ref{Nd})  and calculated
numerically at $x=0.2$.
Dashed curves closest to the corresponding solid ones are obtained
in MFA, using (\ref{eta3}), (\ref{eta4}), (\ref{ss3}) (together
with (\ref{Ni})), respectively.}
\end{SCfigure}

In these terms, one obtains
\begin{equation}
\sigma^\rMFA_\mathrm{BE}=\Lambda_\mathrm{BE}(z,\mu),\quad
\cN^\rMFA_\mathrm{BE}=\frac{z}{1-z}.
\label{eta3}
\end{equation}

Behavior of both (\ref{eta1}) at $d_N=1$ and (\ref{eta3}) in Fig.~1a
justifies applicability of our approach and witnesses on decreasing
effective mass due to gravity and intensifying time dilation with growing
mean number of BHs.

To interpret the results, it needs the account for the admissibility
condition $\cN<\mu^{-3}$ in order to place the $\cN$
black holes (without overlap) within the total volume
$V=4\pi R^3/3>4\pi m^3{\cal N}/3$.

\textbf{Classical Statistics.}
This case of indistinguishable particles uses $d_N=1/N!$ and leads to
the auxiliary functions:
\begin{equation}
\psi_\mathrm{Cl}(z)=\rme^z,\quad
\Lambda_\mathrm{Cl}(z,\mu)=\frac{\Gamma(\mu^{-1})\,\rme^{-z}}{\mu\,\Gamma(\mu^{-1}+1)}
M\left(\mu^{-1},\mu^{-1}+1,z\right),
\end{equation}
where $M(a,b,z)$ is the Kummer's function; $\Gamma(z)$ is the gamma-function.

Using these at arbitrary $z\geq0$, one has immediately
\begin{equation}\label{eta4}
\sigma^\rMFA_\mathrm{Cl}=\Lambda_\mathrm{Cl}(z,\mu),\quad
\cN^\rMFA_\mathrm{Cl}=z.
\end{equation}
We see in Fig.~1b a coincidence of
the numerically calculated dependence $\sigma_\mathrm{Cl}(\cN_\mathrm{Cl})$ with
the analytically derived one $\sigma^\rMFA_\mathrm{Cl}(\cN^\rMFA_\mathrm{Cl})$.
It is easy to observe also a similarity between functions behavior in the BE and Cl cases.

\textbf{Infinite Statistics.}
This type of statistics is meant as a special $q=0$ case of so-called quon statistics
whose defining commutation relation reads $a_ia^\dag_j-qa^\dag_ja_i=\delta_{i,j}$
(the other special values $q=-1$ and $+1$ correspond to the Fermi and Bose cases respectively).
So within infinite or ``Inf'' statistics we have $a_ia^\dag_j=\delta_{i,j}$. More details about
Inf-statistics can be found in \cite{Strom93,Med96,Gr90,Moh90,Ng,Mir13}

Application of this type of statistics to the extremal BHs is motivated by the
work~\cite{Strom93}. The subsequent study of the infinite statistics in \cite{Med96}
leads to a formula for the total number of BHs, which is easily adapted to our model to yield
\begin{equation}\label{Ni}
\cN^\rMFA_\mathrm{Inf}=\frac{1}{z^{-1}-z}
=\frac{1}{2}\left(\frac{z}{1+z}+\frac{z}{1-z}\right),\quad
0<z<1.
\end{equation}
To compute a mean field $\sigma^\rMFA_\mathrm{Inf}=\Lambda_\mathrm{Inf}(z,\mu)$,
we find the weight coefficients $d_N$ and the auxiliary functions by using
the formulas (\ref{Ni}), (\ref{N2}), (\ref{sig}):
\begin{eqnarray}
d_N&=&\frac{\Gamma([N/2]+1/2)}{\Gamma(1/2)\, \Gamma([N/2]+1)};\\
\psi_\mathrm{Inf}(z)&=&\sqrt{\frac{1+z}{1-z}};\\
\Lambda_\mathrm{Inf}(z,\mu)&=&\sqrt{\frac{1-z}{1+z}}\left[
{}_2F_1\left(\frac{1}{2},\frac{1}{2\mu};\frac{1}{2\mu}+1;z^2\right)\right.
\nonumber\\
&&\left.+\frac{z}{1+\mu}\,{}_2F_1\left(\frac{1}{2},\frac{1}{2\mu}+\frac{1}{2};\frac{1}{2\mu}+\frac{3}{2};z^2\right)\right].
\label{ss3}
\end{eqnarray}
Here $[N/2]$ means the integer part of number $N/2$ and ${}_2F_1(a,b;c;z)$
is the Gauss hypergeometric function. Coefficients $d_N$ are defined so
that $d_{2k+1}=d_{2k}$, $k\in\mathbb{N}$.

It is instructive to compare the mean field behavior within the
three statistics admitting unlimited number of particles.
We observe in Fig.~1c that $\sigma_\mathrm{Inf}$
is larger than $\sigma_\mathrm{BE}$ and $\sigma_\mathrm{Cl}$ in
magnitude at the same $\cN$, what is explained by effective
repulsion produced by contribution of the Fermi--Dirac statistics
and described by the term $z/(1+z)$ in (\ref{Ni}). On the other
hand, the fact that the value of $\sigma_\mathrm{Cl}$ is the smallest
one among all, is rather the result of accounting for (indistinguishable)
replicas of multi-BH system which act independently on the test
particle.

Thus, we see that the gravity, growing with increasing BH number,
leads to decreasing the energy-mass of a probe particle and to intensifying
time dilation. However, the (quantum) statistics significantly affects
this tendency that we consider.

\begin{figure}
\begin{center}
\includegraphics[width=8cm]{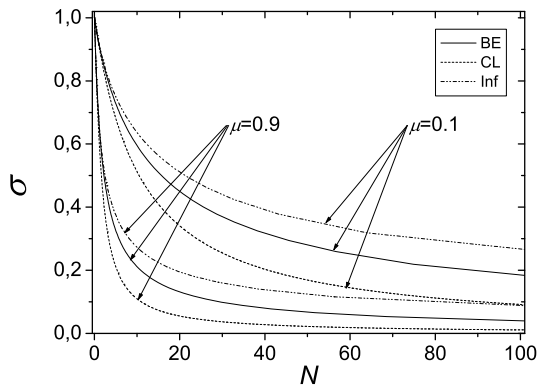}
\end{center}
\vspace*{-5mm}
\caption{\small Mean field $\sigma$ versus the BH mean number $\cN$
at $x=1$ and different values of $\mu$. Curves represent the dependence
of $\sigma_\rms$ on $\cN_\rms$, given parametrically by (\ref{eta1}),
(\ref{Nd})  and calculated numerically within the BE-, Cl- and Inf-statistics.}
\end{figure}

A comparison of the behavior of the mean field $\sigma_\rms$ in these
statistics can be also continued in Fig.~2. However, in this case of
$x=1$, at $m_0=m$ and $T\simeq m$, we can see the relative value of
the effective mass $m_*/m$ of an individual black hole as a result of
the gravitational influence only (without electrostatic) of the rest BHs
from the ensemble.

Finally we would like to consider the case of the Fermi--Dirac statistics
with the limited number of allowed states, which is introduced as follows.

\textbf{Fermi--Dirac Statistics.}
Let $g$ be the number of admissible (energy) states of the BH system
in volume $V$. Therefore, the coefficients $d_N=0$ at $N>g$. The number
of ways of distributing $N\leq g$ indistinguishable particles among
the $g$ energy levels, with a maximum one particle per level, is
given by the binomial coefficient,
\begin{equation}
d_N=\frac{g!}{N!\,(g-N)!}.
\end{equation}
It leads immediately to the auxiliary functions defined at $0\leq z<1$:
\begin{eqnarray}
\psi_\mathrm{FD}(z)&\equiv&\sum\limits_{N=0}^{g} d_N\, z^N=(1+z)^{g},\\
\Lambda_\mathrm{FD}(z,\mu)&\equiv&\frac{1}{\psi_\mathrm{FD}(z)}
\sum\limits_{N=0}^{g} \frac{d_N\,z^N}{1+\mu N}
\nonumber\\
&=&\frac{{}_2F_1(-g,\mu^{-1};1+\mu^{-1};-z)}{(1+z)^{g}}.
\end{eqnarray}
The last function can be related with the Jacobi polynomials $P^{(\alpha,\beta)}_n(x)$.

Using these, one obtains
\begin{equation}
\sigma^\rMFA_\mathrm{FD}=\Lambda_\mathrm{FD}(z,\mu),\quad
\cN^\rMFA_\mathrm{FD}=g\frac{z}{1+z}.
\label{Nfd}
\end{equation}

The limit $g=1$ gives us the maximal energy $m_0/(1+\mu)$ of the test particle in the system
(see (\ref{EN2})), associated with the Fermi level. On the other hand, the whole (homogeneous)
BHs system may be treated effectively as a single particle which is characterized by
the parameter $\mu$. Thus, this situation also indicates that other interactions (like
spin-spin interactions) contribute nothing to energy here.

\begin{figure}
\begin{center}
\includegraphics[width=8cm]{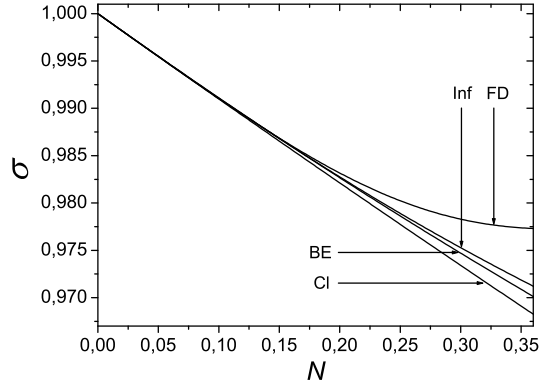}
\end{center}
\vspace*{-5mm}
\caption{\small Mean field $\sigma$ versus the BH mean number $\cN$
at $\mu=0.1$. Curves represent the dependence
of $\sigma_\rms$ on $\cN_\rms$, given parametrically by (\ref{smallN})--(\ref{smallCorr}).}
\end{figure}

Now, let us compare the effect of four statistics (namely, BE,
FD at $g=1$, Inf, and Cl), using the expansion of $\Lambda_\rms(z,\mu)$ and
$\cN^\rMFA_\rms(z)$ over fugacity $z\ll1$,  and setting $m_0=0$ ($x=0$) in final
expressions.
That is, we would like to present the theoretical results which are independent
of the test particle presence at small BH density, requiring simultaneously $\mu\to0$.
Omitting the ``MFA'' abbreviation, we have
\begin{eqnarray}
\cN_\rms(z)&=&z\,[1+\nu_\rms(z)],
\label{smallN}\\
\sigma_\rms(z,\mu)&=&1-\frac{\mu z}{1+\mu}\,[1+\lambda_\rms(z,\mu)],
\label{smallS}
\end{eqnarray}
where the corrections $\nu_\rms(z)$ and $\lambda_\rms(z,\mu)$, defining the deviation, are
\begin{eqnarray}
&&\nu_\mathrm{BE}=z+z^2,\quad
\nu_\mathrm{FD}=-z+z^2,\quad
\nu_\mathrm{Inf}=z^2,\quad
\nu_\mathrm{Cl}\equiv0,\\
&&\lambda_\mathrm{BE}=\frac{z}{1+2\mu},\quad
\lambda_\mathrm{FD}=-z,\quad
\lambda_\mathrm{Inf}=\lambda_\mathrm{Cl}=-\frac{\mu z}{1+2\mu}.
\label{smallCorr}
\end{eqnarray}
We see that $\nu_\mathrm{Inf}=(\nu_\mathrm{BE}+\nu_\mathrm{FD})/2$ and
$\lambda_\mathrm{Inf}=(\lambda_\mathrm{BE}+\lambda_\mathrm{FD})/2$, as
it was mentioned before.

Fig.~3, based on the approximate formulas (\ref{smallN})--(\ref{smallCorr}),
demonstrates the weakening of gravitational effects (equivalently, enhancing
electrostatic repulsion) because of the Pauli principle in the system of
fermion-like extremal BHs.

We should stress that the complete picture of time dilation
within the FD-statistics, which is described by (\ref{Nfd}), shows
significant difference from the BE- and Inf-statistics. First, the mean
number of BHs tends to its limit value equal to infinity for BE- and
Inf-statistics, but to $g/2$ for FD-statistics. Correspondingly, the limit
value of time dilation  $\sigma^\rMFA_\rms$ falls to zero in the BE-
and Inf-statistics, while in the case of FD-statistics $\sigma^\rMFA_\mathrm{FD}$
reaches some finite value, determined by $\mu$.

For this reason, to evaluate the effect of FD-statistics and to compare it with
the BE- and Inf-statistics (all the types treated within approximation
(\ref{smallN})--(\ref{smallCorr}), what is shown in Fig.~3), we replace
the restricted functions (\ref{Nfd}) of $z$ by expressions (\ref{smallN}),
(\ref{smallS}) which are in principle unlimited. In effect, the approximate
results basically differ from the exact ones by construction.

We have already interpreted situation for $g=1$ above. The case $g>1$ demands
taking into account the different states of BHs, which can be associated
both with their positions in space and with internal degrees of freedom
that require detailed study, but not within our model.

\section{Discussion}

Let us first discuss the extremality condition, serving as starting point for
the model that we have described. Imposing the equality of Coulomb
and Newton's forces, we find the relationship between mass $m$ and electric charge
$Z$ (in units of the electron charge):
\begin{eqnarray}
&&\alpha^{-1/2}\frac{m}{M_{\mathrm{Pl}}}=|Z|,\\
&&M_{\mathrm{Pl}}\simeq 2.176\cdot10^{-8}\,\mathrm{kg}\simeq 1.22\cdot10^{19}\,\mathrm{GeV}\cdot c^{-2},\quad
\alpha^{-1/2}\simeq 11.7,
\nonumber
\end{eqnarray}
where $M_{\mathrm{Pl}}=\sqrt{\hbar c/G}$ is the Planck mass;
$\alpha=e^2/(4\pi\varepsilon_0\hbar c)$ is the fine-structure constant.

We are not aware of any empirically known {\it elementary} particles capable of satisfying
this condition. This is one of the arguments to basically distinguish extremal black holes
from elementary particles.
Considering here semiclassical extremal black holes in the ground state, neglecting their
excitations (and description of the internal degrees of freedom) in the low-energy picture,
they should be classically endowed with mass $m\geq1.04\cdot10^{18}\,\mathrm{GeV}/c^{2}$
(to provide $|Z|=1$ at least), the magnitude of which may rather correspond to the theories
of grand unification. At the same time, it is not clear whether the case of $m>M_{\mathrm{Pl}}$,
when the Compton wavelength is less than the gravitational radius $r_+=G m/c^2$, implies
a quantum mechanical meaning.

Another fundamental difference between known particles and BHs is found by referring to
statistics which they obey. While the first ones belong to either bosons or fermions,
the extremal black holes may behave very differently~\cite{Strom93}. Although the latter are
the subject of a miscellaneous theoretical study, they provide the logical possibility of
the existence of entities that are neither fermions nor bosons. This follows from the
assumption that there are internal degrees of freedom inside the horizon that can evolve
(making BHs distinguishable), but the exterior configuration of the BHs remains static.
Then the extremal BHs should also scatter as distinguishable particles in the
quantum picture, when the creation/annihilation operators of their states $l$ and $k$ may obey
the relation $a_ka^\dag_l=\delta_{k,l}$ in any representation~\cite{Strom93}, leading to
the infinite statistics.

Complementing these features of extremal BHs with the justified assumption
that they represent stable and final quantum states obtained as a result of Hawking
radiation~\cite{H75,GS92}, we get physical motivation to study them as special particles
with a nontrivial internal structure, which should be manifested in the statistical description
of the ensemble consisting of a large number of BHs. While abandoning the task of detecting
them in the Universe, we look for the physical manifestations of statistics attributed to
BHs by hands on the basis of logical considerations.

To study the statistical properties of the charged extremal black holes, we used the static
solution of Majumdar--Papapetrou to the Einstein--Maxwell equations~\cite{Maj,Pap,HH72}.
Although it implies dynamical equilibrium (of the Coulomb and Newton's long-range forces)
in the system of $N$ BHs under the extremality conditions $m^2_i/M^2_{\mathrm{Pl}}=\alpha Z^2_i$
($i=\overline{1,N}$), we identify the BH characteristics, $m_i=m$ and $Z_i=Z$, to form an
ensemble of identical and frozen ``particles'' with radius $r_+$.
As noted in \cite{Bon93,Bini07}, such an identification makes the force balance more stable
especially if the presence of a test (and ``extremal'') particle is assumed in the problem.
Indeed, we mainly focus on the time dilation effect experienced by the probe. For this
purpose, we formulate the partition function, based on the Gibbs measure, to find
the statistical characteristics in usual manner. We formally introduce the ``temperature''
of a static system for performing calculations, which drops out from the final expressions.

Since the dynamical equilibrium admits only the absence of overlapping horizons, we use here
a uniform distribution of BHs. This allows us to immediately obtain the mean value of
time dilation in the bulk without additional averaging over the probe position in space.
Carrying out the calculations, we neglected the screening of the electric charge, that is
justified in dilute and homogeneous system when the Debye radius loses its meaning.

We find a time dilation on the base of the effective energy-mass of the test particle,
which is not a universal characteristic here, since it does not take into account
other interactions than the gravitational one. While the time dilation is correctly
determined in the limit of vanishing probe mass, the effective energy-mass may indicate
the gravitational effect of the heavy (extremal) probe on the BH system.

The type of statistics is taken into account here by setting the statistical weight
of the configuration with a fixed number of BHs, which can enhance (weaken) the gravity
effect, leading to varying time dilation. These statistical weights determine the mean
number of black holes and the time dilation in the final formulas, which are obtained in
the mean field approximation and do not depend on a probe characteristics, while
the parameter $\mu=r_+/R$ ($R$ is a fixed radius of the system, similar to what occurs
in the Thomas--Fermi approximation) serves as a measure of the BH influence and distribution
in space. Due to the parameter $\mu$, the probe energy in the system represents an energy
band of finite width and can be described by the $\mu$-deformed numbers, which were already
used in \cite{Jan93,GM12,GM14,GM15,GKKhN,ChGN} Thus, the existence of a band structure turns
out to be important for determining the Fermi level within Fermi--Dirac statistics.

Considering here four cases of Bose--Einstein, Fermi--Dirac, infinite and classical
statistics, we would like to conclude the following. Due to the equal statistical weight
of all the BH configurations within Bose--Einstein statistics, the time dilation effect
(in the bulk) turns out to be more significant than within Fermi--Dirac statistics, dictated
by the Pauli principle, and within infinite statistics, whose effect is intermediate between
those two. Although we have included classical statistics for comparison, one can doubt
its physical applicability because of the assumption of the BHs indistinguishability and
the repeated action of the replicas of a single BH ensemble on a probe what led to
a stronger time dilation.

Since the proper time of observer is evaluated by a product of time dilation $\sigma$
and some global time interval, a difference of several percents of $\sigma$ can significantly enhance and
thus affect the rates of evolutionary processes. Turning to the results obtained within the
considered statistics we note that, although these look rather similar, see Figs.~1 and 2, the overall effects
due to the mentioned factor can gain significant differences.

{\bf Acknowledgments.} The authors are grateful to their colleagues
from the Bogolyubov Institute for Theoretical Physics (BITP) for
valuable discussions of the results presented above. This work was
partially supported by The National Academy of Sciences of Ukraine
(project No. 0117U000237).


\end{document}